\documentclass[12pt,preprint]{aastex} 
\usepackage{amsmath,amssymb,amsfonts} 
\usepackage{natbib} 
\def\cm{{\rm\,cm}}  
  
\def\km{{\rm\,km}}

\def\g{{\rm\,g}}

\def\yr{{\rm\,yr}}

\def\lesssim{\mathrel{\hbox{\rlap{\hbox{\lower4pt\hbox{$\sim$}}}\hbox{$<$}}}} 
\def\gtrsim{\mathrel{\hbox{\rlap{\hbox{\lower4pt\hbox{$\sim$}}}\hbox{$>$}}}}

\shorttitle{Propellers and Frogs} 
\shortauthors{Pan and Chiang} 
\begin{document} 
 
\title{The Propeller and the Frog} 
 
\author{Margaret Pan\altaffilmark{1} and Eugene Chiang\altaffilmark{1,2}} 
 
\altaffiltext{1}{Department of Astronomy, University of California, 
    Berkeley, CA 94720} 
\altaffiltext{2}{Department of Earth and Planetary Science, University of California, 
    Berkeley, CA 94720} 
     
\email{mpan@astro.berkeley.edu}

\begin{abstract} 
 
``Propellers'' in planetary rings are disturbances in ring material
  excited by moonlets that open only partial gaps.  We describe a new
  type of co-orbital resonance that can explain the observed
  non-Keplerian motions of propellers.  The resonance is between the
  moonlet underlying the propeller, and co-orbiting ring particles
  downstream of the moonlet where the gap closes.  The moonlet
  librates within the gap about an equilibrium point established by
  co-orbiting material and stabilized by the Coriolis force. In the
  limit of small libration amplitude, the libration period scales
  linearly with the gap azimuthal width and inversely as the square
  root of the co-orbital mass. The new resonance recalls but is
  distinct from conventional horseshoe and tadpole orbits; we call it
  the ``frog'' resonance, after the relevant term in equine hoof
  anatomy.  For a ring surface density and gap geometry appropriate
  for the propeller Bl\'eriot in Saturn's A~ring, our theory predicts
  a libration period of $\sim$4 years, similar to the $\sim$3.7 year
  period over which Bl\'eriot's orbital longitude is observed to
  vary. These librations should be subtracted from the longitude data
  before any inferences about moonlet migration are made.
\end{abstract} 
 
\section{INTRODUCTION} 
\label{sec:introduction} 
 
Satellites embedded within planetary rings open gaps
\citep{goldreichtremaine80}. Ring particles passing by a satellite are
gravitationally repelled so that the satellite is surrounded by an
underdensity of ring material.  In Saturn's rings, satellites larger
than a few km open gaps extending a full $2\pi$ radians in azimuth
(e.g., Pan in the Encke gap; \citealt{showalteretal86}). For smaller
moonlets, physical collisions and gravitational interactions between
ring particles diffuse particles back into the gap downstream of the
moonlet.  Thus partial gaps are produced whose azimuthal extents
depend on ring viscosity and moonlet mass
\citep{spahn00,srem02,sei05,lewis09}.
 
The {\it Cassini} spacecraft has imaged such partial gaps
(\citealt{tiscarenoetal06}; \citealt{srem07};
\citealt{tiscarenoetal08}; \citealt{tiscareno10}, hereafter
T10). Because of Keplerian shear, density perturbations excited at a
moonlet's position drift toward greater longitudes inside the
moonlet's orbit and toward lesser longitudes outside, forming a pair
of features dubbed a ``propeller'' for its S-like shape. According to
numerical simulations, a propeller's radial width (the radial offset
between the azimuthally extended propeller ``blades'') is $\delta r
\sim 4$ Hill radii of the moonlet \citep{sei05,lewis09}. Combining
this result with measurements of propellers' radial widths from {\it
  Cassini} images, T10 infer moonlet radii of $\sim$0.1--1~km.
 
The appearance of a given propeller varies significantly from image to 
image, depending on the illumination and viewing geometries, the 3D 
structure of the density perturbations, and the optical properties of 
ring particles. These effects have not yet been disentangled.  Still, 
it seems clear that at least for some propellers, the gaps' azimuthal 
(longitudinal) lengths are much larger than their radial widths. Panel 
(e) of Figure~1 of T10 shows an S-shaped bright lobe of radial width 
$\delta r \sim 5.5$ km and azimuthal length $\sim$100~km---this is the 
propeller Bl\'eriot, the largest and most extensively 
imaged---embedded within a still longer dark gap whose azimuthal 
half-length (half the distance between gap ends) is $L_\phi \sim 500 
\km \sim 300$ Hill radii. 
 
Intriguingly, Bl\'eriot displays non-Keplerian motion (T10). Over several 
years, Bl\'eriot's orbital longitude deviated from a strictly 
Keplerian solution, showing residuals that varied nearly sinusoidally 
in time with a period of $\sim$3.7 yr and an amplitude (half the 
distance from peak to trough) of $\sim$0.1 deg or $\sim$200 km, an 
amount less than but of order $L_\phi$. Data for other propellers are 
much more limited than for Bl\'eriot but suggest similar behavior 
(T10). 
 
The longitude variations imply semimajor axis variations of the
underlying moonlet. \citet{cridaetal} and \citet{rein10} investigated
semimajor axis variations due to stochastic torques exerted by
self-gravitating wakes of ring particles \citep{salo95}. Over
timescales of years, stochastic torques might give rise to observable
drifts in longitude, as shown in Figure 6 of \citet{rein10}.  But
the observed 3.7-year period of the longitude variations does not
arise naturally from stochastic torques, which are driven by wakes
that have lifetimes of order one orbit period $\simeq$
14~hours.\footnote{\citet{cridaetal} and \citet{rein10} completed
  their studies before Figure 4 of T10 was made public.}
Alternatively, Bl\'eriot's moonlet may librate within a resonance
established by another larger moon (T10). However, no such resonant
partner has been identified. Furthermore, many other propellers show
longitude deviations of similar magnitude to Bl\'eriot's, and invoking
a separate partner moon for every propeller seems unnatural.
 
Here we propose that Bl\'eriot and propellers like it are indeed 
participating in resonances. But the resonances are with nearby ring 
material, not with other moons. Each propeller's moonlet librates 
within the potential established by co-orbiting ring particles at the 
ends of the long gaps, at distances $L_\phi$ away.  This new type of 
co-orbital resonance, reminiscent of that between the Saturnian 
satellites Janus and Epimetheus \citep{yoder83}, shares properties 
with both tadpole and horseshoe orbits in the restricted three-body 
problem (see, for example, \citealt{murray99}). 
 
\section{THE FROG RESONANCE} 
\label{sec:tadpole} 
\begin{figure} 
\includegraphics[angle=270,scale=.63]{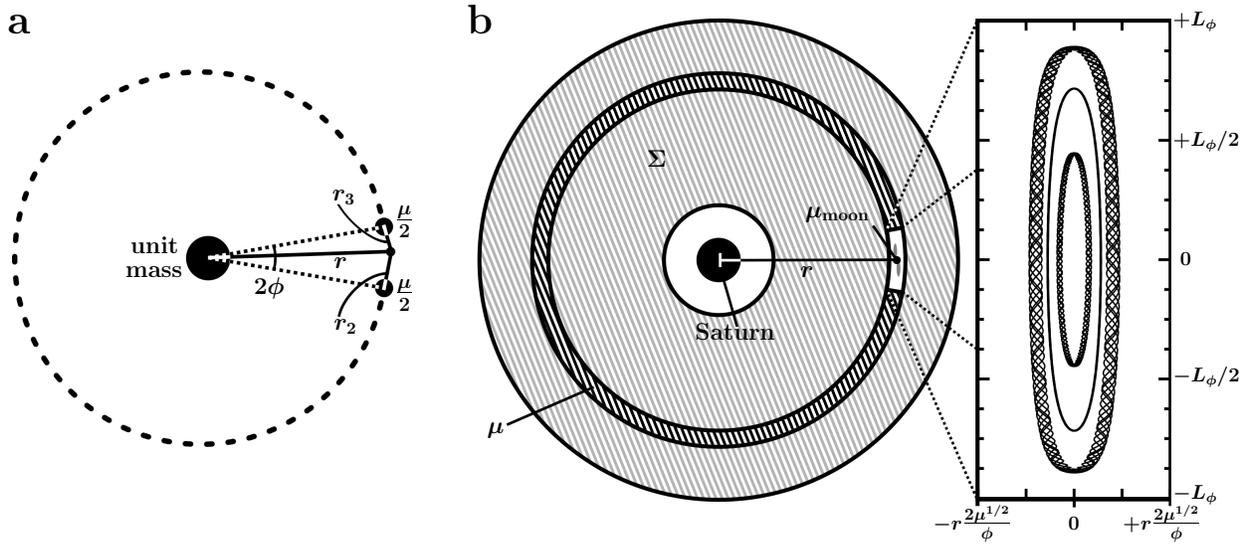} 
\caption{Schematic of our dynamical model for Saturnian 
  propellers. Panel (a) depicts our toy model: a test-particle moonlet 
  moving between two co-orbiting point-mass secondaries. All lengths 
  and angles are evaluated in the frame co-rotating with the 
  secondaries.  Panel (b) illustrates how the propeller---the S-shaped 
  figure centered on the moonlet---is embedded in a partial gap within 
  Saturn's rings.  The point masses in panel (a) represent the ends of 
  the heavily shaded, horseshoe-shaped, co-orbital ring in panel 
  (b). The zoomed-in panel on the right shows sample trajectories of a 
  test particle computed with the co-orbital mass distributed 
  uniformly in azimuth outside the gap (see also Figure 
  \ref{fig:smooth}b) using $\mu = 10^{-4}$ and $L_\phi/r = 0.4$.  Fast 
  epicyclic motion of different amplitudes is visible on top of the 
  azimuthally elongated ``frog'' orbit of the guiding center. 
  \label{fig:schematic}} 
\end{figure}

We begin with a toy model of the co-orbital material 
(Figure~\ref{fig:schematic}a): two identical point masses 
(``secondaries'') each of mass $\mu/2 \ll 1$ orbiting a central body 
(``primary'') of mass 1. The secondaries reside on circular, 
co-planar, Keplerian orbits of radius 1 and angular frequency 1. The 
azimuthal separation of the secondaries is $2\phi \ll 2\pi$.  The 
secondaries are meant to represent ring material at the ends of the 
gap opened by a propeller-moonlet. 
 
The secondaries establish an equilibrium point between them at azimuth 
$\theta = 0$ and radius $r\simeq 1$ (in cylindrical coordinates in the 
frame co-rotating with the secondaries). Just as for the triangular 
Lagrange points, this equilibrium point is dynamically stable because 
of the Coriolis force.  And as with conventional tadpole and horseshoe 
orbits, test particle motions about our equilibrium point decompose 
into a fast epicycle of period $2\pi$ and a slow libration of the 
guiding center (semimajor axis). We propose that the slow libration is 
the non-Keplerian motion reported by T10. 
 
We now derive the libration period and the shape of the guiding center 
orbit in the limit of small libration amplitude.  In the frame 
rotating with angular frequency 1, the equations of motion for the 
(assumed massless) moonlet read 
\begin{eqnarray} 
{\ddot r}-r{\dot\theta}^2-2r{\dot\theta} & = & \frac{\partial U}{\partial r}  \label{eqn:r}\\ 
r{\ddot\theta}+2{\dot r}{\dot\theta} +2{\dot r} & = & \frac{1}{r}\frac{\partial  
U}{\partial\theta} \label{eqn:theta} 
\end{eqnarray} 
where $U$ is the celestial mechanician's potential, 
\begin{equation} 
U = \frac{1}{r} + \frac{\mu}{2r_2} + \frac{\mu}{2r_3} + \frac{1}{2}r^2 \,\, , \label{eqn:u} 
\end{equation} 
and the distances between the moonlet and the secondaries are 
\begin{equation} 
r_2 = \sqrt{1+r^2-2r\cos(\theta+\phi)} \;\;\; , \;\;\; 
r_3 = \sqrt{1+r^2-2r\cos(\theta-\phi)} \;\;\; . 
\end{equation} 
Note that we are ignoring the displacement of the center-of-mass from 
the primary. As shown below, the associated indirect term of the 
potential, though crucial for the stability of conventional tadpoles, 
is not essential for our problem. Moreover, in the more realistic 
situation where the mass in the secondaries is spread smoothly over 
the co-orbital region---i.e., over nearly the full $2\pi$ rad in 
azimuth, excepting the gap of width $2\phi$---the barycentric 
displacement is much smaller than in the present two-point-mass 
problem. We will consider this more realistic case at the end of this 
section. 
 
\begin{figure} 
\epsscale{1} 
\plotone{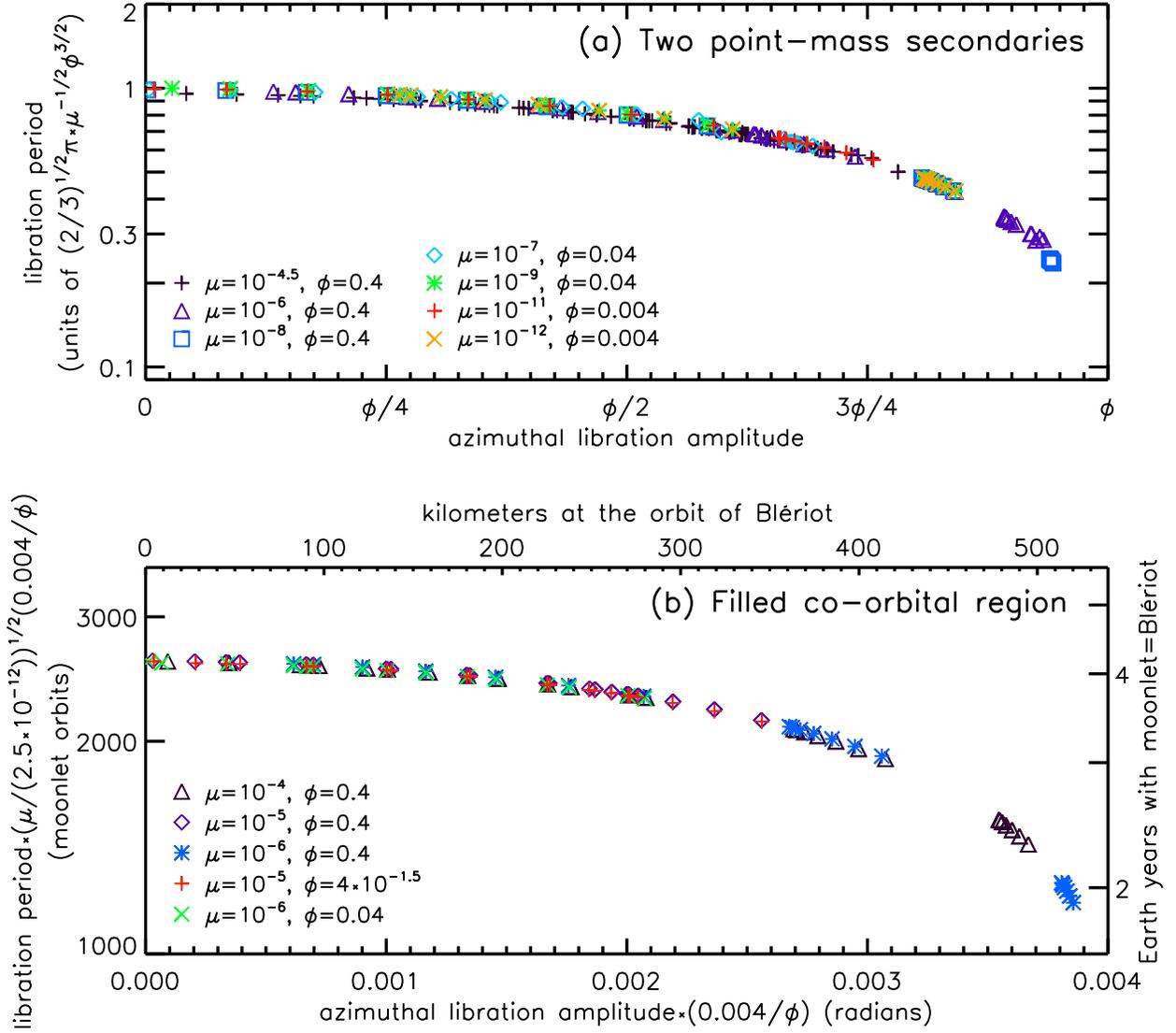} 
\caption{Libration period of a test particle as a function of 
  azimuthal libration amplitude, computed from direct numerical 
  integrations for a variety of co-orbital masses $\mu$ and gap 
  azimuthal half-widths $\phi$. (a) Results for the toy model with two 
  point-mass secondaries. The y-axis is scaled to $\mu$ and $\phi$ 
  according to equation (\ref{eqn:period}), whose validity is 
  confirmed by the data for small libration amplitude. (b) Results for the 
  smoothed-out case where the co-orbital mass $\mu$ is modeled as a 
  line source of uniform mass density extending in azimuth from $\phi$ 
  to $2\pi-\phi$ (passing through $\pi$; see Figure 
  \ref{fig:schematic}).  The y-axis is scaled according to equation 
  (\ref{eqn:period_smooth}); that all data fall on the same curve 
  confirms the scalings of this equation. Note that in both the 
  two-point-mass and smoothed-out models, libration periods decrease 
  as the libration amplitude increases. This behavior is similar to 
  that of conventional horseshoe orbits.} 
\label{fig:smooth} 
\end{figure}

We expand $U$ in the limit of small displacements from the equilibrium 
point, $\Delta \equiv r - 1 \ll 1$ and $\theta < \phi \ll \pi$: 
\begin{equation} 
U \approx \frac{3}{2} + \frac{3}{2} \Delta^2 + \frac{\mu}{\phi} \left( 1 - \frac{\Delta}{2} - \frac{\Delta^2}{2\phi^2} + \frac{\theta^2}{\phi^2} \right) \,. \label{eqn:approxu} 
\end{equation} 
We insert (\ref{eqn:approxu}) into (\ref{eqn:r})--(\ref{eqn:theta}), 
taking $d/dt \ll 1$ and keeping only leading-order terms to 
filter out the fast epicyclic motion. We also assume that $\phi \gg 
\mu^{1/3}$, i.e., we assume that the gap length is much larger than 
the Hill sphere of the secondary (not to be confused with the Hill 
sphere of the moonlet). The fixed point is then  
\begin{equation} 
\Delta\simeq\frac{\mu}{6\phi} \;\;\; , \;\;\; \theta=0 \label{eqn:fixedpt} 
\end{equation} 
and the equations of motion about it are 
\begin{eqnarray} 
{\dot \theta} & = & -3\Delta/2 \label{eqn:leadr}\\ 
{\ddot \theta} + 2 {\dot \Delta} & = & (2\mu/\phi^3)\, \theta \,. \label{eqn:leadtheta} 
\end{eqnarray} 
Equation (\ref{eqn:leadr}) states that the test particle moves 
according to the Kepler shear. Taking the time derivative of 
(\ref{eqn:leadr}) and inserting the result into (\ref{eqn:leadtheta}), 
we have 
\begin{equation} 
{\ddot \theta} = - (6\mu/\phi^3)\, \theta 
\end{equation} 
which yields harmonic motion of period 
\begin{equation} 
P_{\rm lib} = \frac{\pi\sqrt{2}}{\sqrt{3}} \, \frac{\phi^{3/2}}{\mu^{1/2}} \,\,\,\,\,\, \text{for two point-mass secondaries}  
\label{eqn:period} 
\end{equation} 
in units where $2\pi$ is the local orbital period. 
Since $\Delta$ is largest when ${\dot \theta}$ is largest, the aspect ratio 
of the guiding center orbit is 
\begin{equation} 
\frac{\max \Delta}{\max \theta} = \frac{4}{\sqrt{6}} \frac{\mu^{1/2}}{\phi^{3/2}} \,\,\,\,\,\, \text{for two point-mass secondaries.}  
\label{eqn:aspect} 
\end{equation} 
Numerical integrations of the two point-mass case agree well with 
equations (\ref{eqn:period}) (Figure~\ref{fig:smooth}a) and 
(\ref{eqn:aspect}) (data not shown). 
 
A few comments: 
\begin{enumerate} 
\item{As with conventional horseshoe orbits but not 
conventional tadpoles, stable librations here do not require the 
indirect potential.} 
\item{As shown in Figure \ref{fig:smooth}a, the libration period 
    decreases with increasing libration amplitude; again, this 
    behavior resembles that of horseshoes but not of tadpoles.} 
\item{Without the Coriolis force term 
$2{\dot\Delta}$ in equation (\ref{eqn:leadtheta}), stable librations 
would not occur. Thus the Coriolis force acts as a restoring force, 
just as it does for tadpoles.} 
\item{Like the triangular Lagrange points, our fixed point corresponds 
  to a local minimum of $U$ and a local maximum of the physical 
  potential $-U$. If the energy of a particle in resonance is 
  dissipated slowly compared to the libration period, the libration 
  amplitude should grow with time, just as for conventional 
  tadpoles. Our numerical simulations confirm this.} 
\item{Given $\phi \gg 
\mu^{1/3}$, the guiding center orbit is thinner in $r$ than in 
$\theta$.  Conventional tadpole orbits are even thinner: $(\max \Delta 
/ \max \theta)_{\rm tadpole} = \sqrt{3\mu}$.} 
\end{enumerate} 
Given these commonalities with tadpoles and horseshoes, we refer to 
our guiding center orbits as ``frog orbits.''\footnote{The rearward 
  portion of a horse's hoof---what would lie in the space between the 
  ends of a horseshoe---is called a frog. See, e.g., \hfill\\
  \url{http://www.horsemanmagazine.com/wp-content/uploads/2008/09/horse-hoof-frog.jpg}.} 
 
We now address the more realistic situation where the secondary masses 
are spread uniformly in azimuth to fill the region $\theta = \phi$ to 
$\theta = 2\pi - \phi$ (passing through $\pi$).  In this case we might 
expect only the mass within an angle $\sim$$\phi$ of either end of the 
gap to influence the moonlet. In other words, the effective mass of 
the co-orbital region may be contained at azimuths $|\theta|$ between 
$\phi$ and $\sim$$2\phi$. If so, then the scalings in equations 
(\ref{eqn:period}) and (\ref{eqn:aspect}) still apply with $\mu$ 
replaced by the effective mass $\sim$$(\phi/\pi)\cdot\mu$, where now 
$\mu$ is the mass of the entire ring co-orbiting with the moonlet. Our 
numerical simulations 
(Figures~\ref{fig:schematic}b,~\ref{fig:smooth}b) confirm the 
resulting scalings of period and aspect ratio with $\mu$ and $\phi$ 
and yield coefficients such that 
\begin{eqnarray} 
P_{\rm lib}& \simeq& 6.4 \, \frac{\phi}{\mu^{1/2}}  
\label{eqn:period_smooth}\\ 
\frac{\max \Delta}{\max \theta}& \simeq& 0.7 \, \frac{\mu^{1/2}}{\phi} \,\,\,\,\,  
\label{eqn:aspect_smooth} 
\text{for $\mu \equiv$ mass of entire co-orbital ring} 
\end{eqnarray} 
in the limit of small libration amplitude.  For parameters inspired by 
Bl\'eriot and its environment---i.e., for $\phi = L_{\phi}/r \sim 500 
\km/135000 \km \sim 0.004$ and $\mu \sim 2\pi \Sigma r \,\delta r / 
M_{\rm Saturn} \sim 2.5\times 10^{-12}$, where $\Sigma \sim 
30\,\g\cm^{-2}$ is the disk surface density \citep{colwell09} and 
$\delta r \sim 5.5\,\km$---our numerical integrations yield $P_{\rm 
  lib} \sim$ 4 yr. This is encouragingly close to the $\sim$3.7 yr 
period measured by T10. Note that these parameters satisfy the 
assumptions $\mu\ll 1$, $\mu^{1/3}\ll\phi\ll 1$, and $\Delta<\delta 
r\ll1$ that we made to derive 
equations~(\ref{eqn:period})--(\ref{eqn:aspect_smooth}). Also, 
assuming a moonlet density of $\sim$1~g/cc and the median moonlet 
radius $\sim$0.75~km inferred by T10, the moonlet-to-Saturn mass ratio 
$\mu_\mathrm{moon}$ is $\sim$$3\times 10^{-15}\ll\mu$ as required by 
our assumption that the moonlet mass is negligible. 
 
\section{SUMMARY AND DISCUSSION} 
\label{sec:discuss} 
 
In this work we identified a new kind of orbital resonance between a 
moonlet embedded within an underdense gap of angular extent $\phi\ll 
\pi$ and disk material co-orbiting with the moonlet beyond the 
ends of the gap. We found formulae for the resonant libration period 
$P_{\rm lib}$ (equation \ref{eqn:period_smooth}) and the 
radial/azimuthal aspect ratio of the guiding center orbit (equation 
\ref{eqn:aspect_smooth}) in terms of $\phi$ and the co-orbital disk 
mass $\mu$. These librations resemble standard tadpole orbits because 
$P_{\rm lib} \propto 1/\mu^{1/2}$ and because the Coriolis force 
stabilizes the resonance. But whereas tadpoles also require the 
indirect potential for stability, the librations discussed here do 
not and are more like horseshoes in this respect. We propose that our 
new ``frog resonance'' explains the non-Keplerian motion seen for 
Saturnian propeller features such as Bl\'{e}riot. The observed 
timescale over which Bl\'eriot's orbital longitude residuals vary 
($\sim$3.7 yr) is well reproduced by our theory.  Measurements of 
non-Keplerian motion and their interpretation as resonant frog 
librations thus offer new constraints on ring surface density (through 
$\mu$) and gap geometry (through $\phi$) independent of radiative 
transfer models of scattered light images. 
 
Our simple resonance model does not include interactions with other 
satellites, nor does it account for disk torques responsible for Type 
I or Type II migration (for migration in the context
of gas disks, see \citealt{ward97}; analogous effects
for particle disks like Saturn's rings are discussed
by, e.g., \citealt{cridaetal}).  Since no propellers are seen 
at the locations of strong mean-motion resonances with the major 
Saturnian moons, and since Bl\'{e}riot at least has no obvious 
mean-motion resonant partner, we believe it reasonable to neglect such 
resonances. Secular interactions do not alter the semimajor axis of a 
propeller-moonlet; moreover, the eccentricities involved are too small 
for secular precession to be relevant. 
 
Type I migration is driven by imbalanced Lindblad torques from
the disks exterior and interior to the moonlet's orbit.
The torque from each side is dominated by material at the edge
of the gap, located at a radial distance $\pm \delta r$ away from the moonlet.
If we assume that the torque from the outer disk exceeds
that from the inner disk by of order $\delta r/ r$, where $r$
is the moonlet's orbital radius, then the Type I migration rate is given by
\begin{equation} 
\dot{r}_{\rm Type \,I} \sim -\mu_{\rm disk} \cdot 
\mu_{\rm moon} \cdot 
\left(\frac{r}{\delta r}\right)^2 \cdot 
v_{\rm Kepler} 
\sim -1\,\cm/\yr,  
\label{eqn:typei} 
\end{equation} 
where $\mu_{\rm disk} = \Sigma r^2 / M_{\rm Saturn}$; $\Sigma$ is the
unperturbed local surface density; $\mu_{\rm moon}$ is the mass ratio
of the moonlet to Saturn; and $v_{\rm Kepler}$ is the Keplerian
orbital velocity of the moonlet. Equation~(\ref{eqn:typei}) follows
from the standard impulse approximation (e.g.,
\citealt{dermott84}). Because the gap radial width $\delta r \sim
\mu_{\rm moon}^{1/3} r$ for propellers, the scalings in
equation~(\ref{eqn:typei}) are equivalent to those of equation (39) of
\citet{cridaetal}: $\dot{r}_\mathrm{Type\, I}\propto
\mu_\mathrm{disk}\,\mu_\mathrm{moon}^{1/3}\,v_\mathrm{Kepler}$.  Our
numerical estimate for $\dot{r}_\mathrm{Type\, I}$ in equation
(\ref{eqn:typei}) is made for parameters appropriate to Bl\'{e}riot
and is much too small to contribute significantly to that propeller's
longitude residuals, as those residuals imply an rms radial speed of
$\sim$100 m/yr.  Moreover the Type I drift is of one sign, whereas the
radial drift of Bl\'eriot implied by the observations switches sign.
 
However, since $\delta r / r \sim 4 \times 10^{-5}$, the near
cancellation of Lindblad torques assumed above is delicate and may be
overwhelmed by radial surface density gradients. Any long-term trend
in the longitude residuals, over and above the sinusoidal frog
oscillations which we have analyzed, may therefore help constrain the
radial density profile. For example, Figure 4a of \citet{tiscareno10}
shows an apparent monotonic drift in Bl\'eriot's longitude residuals
of about $+0\fdg 05$ in 3.7 years. If this drift is real and if it
grows quadratically with time---and these are highly uncertain
prospects, as any characterization of a long-term trend depends on the
method used to derive Bl\'eriot's mean motion (M.~Tiscareno, personal
communication)---then such a drift would correspond to an inward
migration speed of $\sim$3 m/yr. The disk just outside Bl\'eriot's
orbit would have a greater surface density than that inside by of
order 1 part in 100. Alternatively, a long-term nonlinear trend in the
longitude could arise from the stochastic torques studied by
\cite{cridaetal} and \cite{rein10}.
 
We have interpreted the non-Keplerian motions of propeller-moonlets as
backreactions of their perturbed disks on the moonlets. Our model,
however, is not self-consistent because we have not considered how the
motions of the moonlet feed back into shaping the gap. We have only
computed the motion of a moonlet, idealized as a test particle,
embedded in a gap whose structure is presumed stationary in the
co-rotating frame. We do not have a theory that predicts the libration
amplitude relative to the gap ends. Energy dissipation (for
example, by interparticle collisions) tends to increase the libration
amplitude, but we do not understand how the amplitude might be damped.
Observationally, whether the gap itself shows non-Keplerian motion, or
equivalently whether the moonlet moves relative to the gap ends, is
difficult to determine (M.~Tiscareno, personal communication). The
appearance of a gap varies from image to image in ways not yet
completely understood, and even measurements of basic quantities such
as a gap's angular size $\phi$ are complicated by azimuthal structure
inside the gap (i.e., gap ends are not simple step functions).
Improvements in radiative transfer modeling of images; more
astrometric measurements of propellers and their relative positions
inside their gaps; and a self-consistent theoretical treatment of how
moonlets force disks and disks force moonlets may yield further
insights.
 
\acknowledgments We thank Matt Tiscareno for sharing his discovery in
advance of publication that propellers exhibit non-Keplerian motion,
and for subsequent discussions.  MP appreciates the hospitality of the
Canadian Institute for Theoretical Astrophysics, where part of the
writing was done. We thank Glen Stewart for a helpful referee's
report, and Joe Burns for encouraging remarks.  We thank Aur\'elien
Crida, John Papaloizou, and Hanno Rein for feedback on our manuscript.
This research was supported in part by the National Science Foundation
and UC Berkeley's Center for Integrative Planetary Science.
 

\end{document}